# Evolution of surface morphology from Stranski–Krastanov growth mode to step-flow growth mode in InSbBi thin films

Chandima Kasun Edirisinghe[1], Anuradha Wijesinghe[1], Anjali Rathore[1], Pradip Adhikari[1], Christopher Rouleau[2], Joon Sue Lee[1,*]

[1]Department of Physics and Astronomy, University of Tennessee, Knoxville, TN 37996, USA

[2]Center for Nanophase Materials Sciences division, Oak Ridge National Laboratory, USA

[*]Corresponding author e-mail: jslee@utk.edu

## Abstract

The incorporation of dilute concentrations of bismuth (Bi) into traditional III-V alloys leads to a significant reduction in bandgap energy, making InSbBi a promising candidate for long-wavelength (<0.17 eV) infrared photosensors. InSbBi offers potential as a platform for exploring spin dynamics and quantum phenomena due to its strong spin-orbit coupling. Despite this promise, the material quality of InSbBi alloys remains inferior to that of conventional III–V semiconductors, primarily due to the substantial challenges associated with incorporating Bi into InSb and producing high-quality InSbBi with varying Bi compositions. In this study, we address these issues by developing a molecular beam epitaxy approach to grow smooth InSbBi thin films with tunable Bi incorporation up to 1.8 atom % through dynamic adjustment of Sb flux and careful control of the interplay between growth temperature and Bi flux. This work paves the way for the development of high-quality InSbBi thin films for applications in photo detection, spintronics, and related quantum technologies.

# I. INTRODUCTION

In recent years, extensive research has focused on III-V semiconductor alloys due to their potential applications in near- and mid-infrared optoelectronic devices. One example is InSb, which has been used extensively in mid-infrared photodetectors.[1,2] However, its energy bandgap of 0.17 eV limits its ability to detect photons in the long-wavelength infrared (LWIR) range.[1] Hence, several III-V ternary alloys such as InTlSb[3,4], InAsBi[5,6], InTlP[7,8], and InTlAs[9,10] have been investigated as suitable candidates. Despite the encouraging results, further reduction of bandgap energy is necessary for high-performance detector operation.[11] InSb has the smallest bandgap in established III-V semiconductors[12], and by incorporating bismuth (Bi), the lower bandgap ternary alloy InSbBi can be formed.[13,14] Hence, InSbBi has become an attractive material for LWIR applications, and recent studies have demonstrated that InSbBi epitaxy films grown by molecular beam epitaxy (MBE) on InSb substrates have demonstrated a reduction of 35 meV per atom % Bi at room temperature as the Bi incorporation increases from 0.6 to 2.4 atom %.[1,13]

The addition of Bi also enhances the spin-orbit splitting energy of III-V semiconductors.[15] This is primarily due to Bi's high atomic number and strong nuclear charge.[16-18] The strong spin-orbit splitting energy, particularly in InSbBi, will enhance the Landau g-factor by altering the energy levels and spin states of electrons[19,20], which may provide exciting opportunities for advanced spintronics applications[21] and quantum phenomena such as topological superconductivity.[22-25]

Theoretical and experimental studies of a significant band gap reduction and bowing of spin-orbit splitting energy in GaAsBi have been reported.[15,18,26-28] However, there remains a limitation in the experimental research on InSbBi, which can be attributed to significant challenges associated with synthesizing high-quality InSbBi samples.[1] Even among the limited studies available, results have often been contradictory; for instance, some researchers have reported contraction of the lattice after Bi incorporation into InSb[11,29] while others[13,14,30] have demonstrated lattice expansion. Furthermore,

Rajpalke *et al.*[13] reported that the growth temperature must be lower than 225°C for Bi incorporation above 1 atom % whereas White *et al.*[14] achieved 1.9% even at a higher growth temperature of 300°C, highlighting a discrepancy in the relationship between Bi incorporation and growth temperature.

In this work, we present a platform that enables precise control over Bi incorporation in epitaxial InSbBi films by focusing on the evolution of the surface morphology from Stranski‑Krastanov (SK) to step-flow growth modes while maintaining a smooth surface morphology without Bi droplets on the surface. This is achieved by careful flux control of antimony (Sb) and Indium (In) and the interplay between growth temperature and Bi flux. The findings of this study will contribute to the fundamental understanding of thin-film technology and offer a practical approach for growing smooth InSbBi films with tunable Bi incorporation, facilitating their integration into hybrid heterostructures for optoelectronic and quantum devices.

## II. EXPERIMENTAL

InSbBi thin films were grown on a homoepitaxial InSb buffer layer on InSb(001) substrates using MBE. 1 cm × 1 cm sized InSb(001) substrates, Ga-bonded on tantalum substrate holders, were outgassed at a manipulator temperature of 350°C for 1 hour in an ultrahigh vacuum (UHV) chamber with a base pressure of $\sim 10^{-10}$ Torr. The substrates were transferred to an interconnected III-V MBE chamber, and the native oxide was thermally desorbed at a substrate temperature of 445°C in the presence of $Sb_2$ flux. Then, an InSb homoepitaxial layer of 100 nm was grown at 400°C. The resulting buffer layer exhibited a streaky (1 × 3) reflection high energy electron diffraction (RHEED) pattern, providing *in-situ* confirmation of the reconstructed InSb surface achieved under Sb-rich growth conditions[31]. Subsequently, InSbBi thin films of 300 nm were (FIG. 1(a)) grown with a systematic variation of growth parameters. Initially, the V/III flux ratio was varied by adjusting the Sb flux dynamically – i.e., changing the position of the Sb valved cracker to provide only the necessary amount of Sb flux to change the RHEED surface reconstruction from (4 × 2) to (1 × 3) while keeping the In and Bi fluxes constant at $4.9\times10^{-7}$ and $1.7\times10^{-8}$ Torr, respectively. We note that the V/III flux ratio must be maintained around unity to grow high-quality InSbBi thin films. A ratio above unity results in a (1 × 3) surface reconstruction while a ratio below unity results in a (4 × 2) surface reconstruction.[31] Following this, the Bi flux was systematically varied from $1.2\times10^{-8}$ to $5.0\times10^{-8}$ Torr, with the V/III flux ratio maintained at unity and the growth temperature kept constant at 200°C. Finally, the growth temperature was varied while keeping the Bi flux constant at $1.7\times10^{-8}$ Torr and the V/III flux ratio at unity. Growth temperatures were determined by using a pyrometer down to 300°C, thereby enabling manipulator thermocouple readings to be estimated down to 200°C.

Crystal structures of InSbBi films were characterized by x-ray diffraction (XRD) using a Malvern Panalytical X'pert3 Materials Research Diffractometer with Cu-kα radiation source (λ=1.54 Å). The dynamical diffraction curves used to calculate the Bi incorporation were generated using the X'Pert Epitaxy software. The InBi lattice constant was chosen to be 6.686 Å with a zinc blende structure for these simulations.[32] A Poisson ratio of 0.3503 was used for simulations.[1,2] Surface morphology of the grown InSbBi films was characterized by atomic force microscopy (AFM) using a Cypher S AFM. Two measurements were carried out for each sample covering a wider area of 10 × 10 $\mu m^2$ and a smaller area of 1 × 1 $\mu m^2$.

## III. RESULTS AND DISCUSSION

### A. Effect of V/III ratio for the growth of InSbBi thin films

A V/III ratio greater than unity leads to accumulation of Bi droplets on the surface as Bi atoms are unable to incorporate substitutionally[1], and Bi-adsorbed InSb under Sb-rich conditions show energetically favored reconstructions stabilized by Sb, despite the existence of Bi atoms.[33,34] To examine the impact of the V/III ratio greater than unity, an InSb film was grown with a V/III ratio of 1.1 at a growth temperature of 200°C, with a Bi flux of $1.7\times10^{-8}$ Torr. The resulting surface morphology shown in the AFM images in Fig. 1(b) reveals the formation of Bi droplets on the surface (left panel). The magnified 1 × 1 μm$^2$ area AFM scan in Fig. 1(b) right panel further highlights the formation of nanostructures oriented along the [110] direction likely formed by Bi.

To examine the effect of a flux ratio at unity, another InSb film was grown with a flux ratio of unity at a growth temperature of 200°C and Bi flux of $1.7\times10^{-8}$ Torr. The resulting surface morphology exhibits a transition from nanostructures to mound formation, as shown in Fig. 1(c), indicating a shift in the growth mode. This morphological evolution will be further discussed in Sec. B: *Change in Structural Morphology with Varying Bi Flux.*

If the V/III ratio is less than unity, Bi incorporation will be suppressed despite the existence of Bi atoms on the growing surface. The strength of the In–Sb bond is considered to be stronger than that of In–Bi. Therefore, the Sb atoms are preferentially incorporated while the Bi atoms are not. Furthermore, the In–Bi bond is easily replaced by the In–Sb bond.[35] This will lead to having rough surfaces. To examine the impact of having a V/III ratio less than unity, a sample was grown with a V/III ratio of 0.9 at a growth temperature of 200°C with a Bi flux of $1.7\times10^{-8}$ Torr, and the resulting surface morphology is shown in the AFM images in Fig. 1(d). Formation of droplets on the surface is visible as shown in Fig. 1(d) left panel. These droplets are likely to be In droplets since the film was grown under In-rich conditions. When zoomed in on the surface, step-flow growth is visible as shown in Fig. 1(d) right panel. Step-flow growth happens due to the growth of InSb, as shown in the AFM images in Fig. 2(a), when Bi is not incorporated into the growing film.

Wide area θ/2θ XRD scans were conducted to investigate the effect of the flux ratio, as shown in Fig. 1(e). In addition to the InSb(00n) substrate peaks, no other peaks corresponding to elemental In or Bi were observed. To further assess the impact of the flux ratio on Bi incorporation, high-resolution θ/2θ XRD scans were performed, as shown in Fig. 1(f). A distinct InSbBi peak is observed when the flux ratio is at unity. However, no InSbBi peaks are detected when the V/III flux ratio is either greater than or less than unity (i.e., 1.1 and 0.9). These results demonstrate that the V/III ratio plays a critical role in the incorporation of Bi into InSb to form InSbBi thin films.

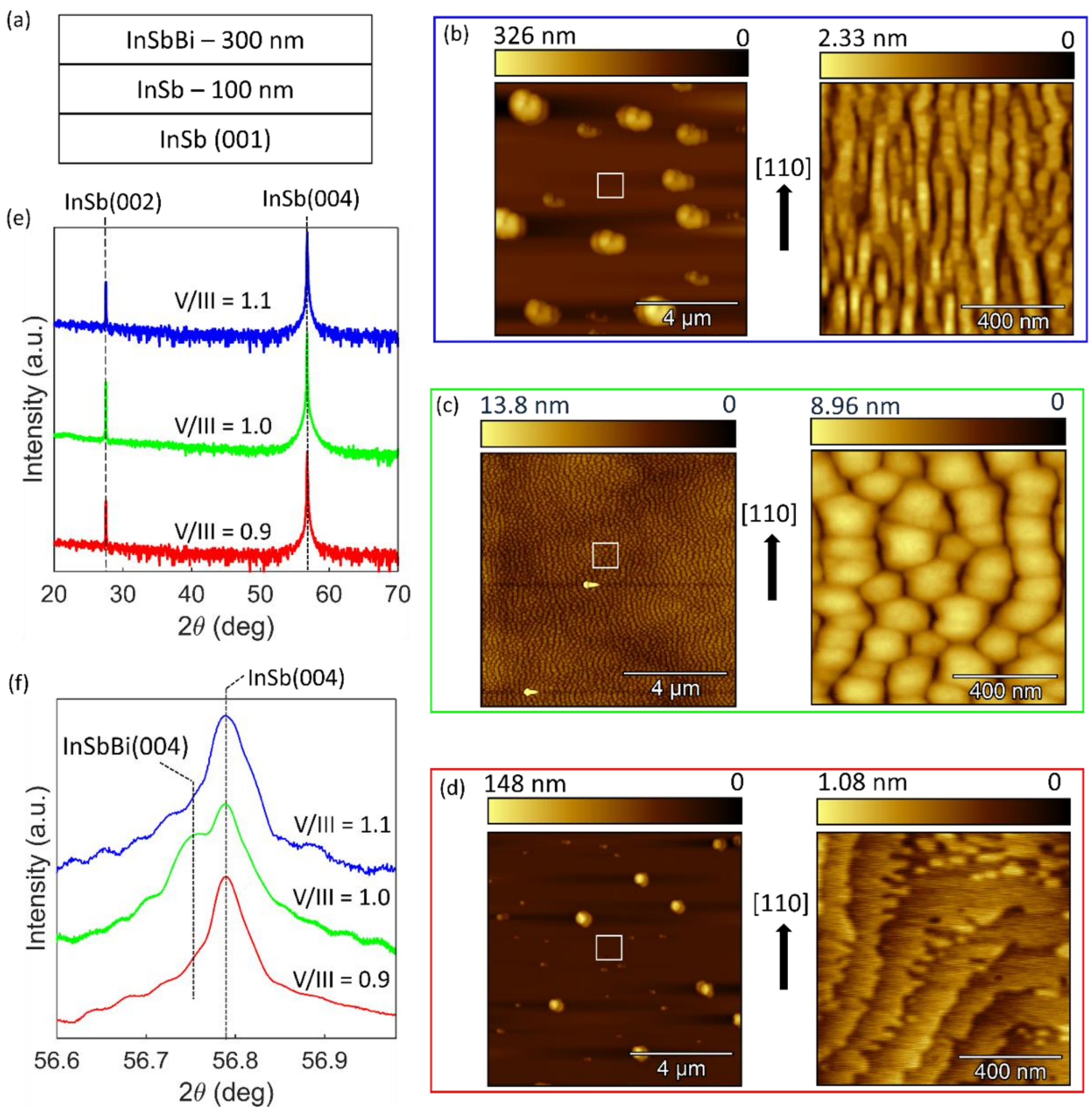

FIG. 1: Effect of V/III ratio on the InSbBi growth. (a) Structure of the samples. (b-d) AFM scans when the flux ratio is 1.1, 1.0, and 0.9, respectively. Left and right panels of (b-d) show surface areas of 10 × 10 μm$^2$ and 1 × 1 μm$^2$, respectively. The magnified regions in the right panels correspond to the areas outlined by the white boxes in the left panels. The black arrow corresponds to the [110] direction of the AFM images. (e) Wide area θ/2θ XRD scans for the samples. (f) Small area θ/2θ XRD scans for the samples.

## B. Change in surface morphology with varying Bi flux

The change in the structural morphology of InSbBi films was studied systematically by varying the Bi flux up to $5.0\times10^{-8}$ Torr while keeping the growth temperature constant at 200°C. The surface morphology changes as the Bi flux varies, as shown in the AFM images in Fig. 2. InSb exhibits step edges (Fig. 2(a)), and with the introduction of Bi flux, the step edges of InSb disappear, leading to a nonequilibrium morphological evolution into "mound" structures, as shown in Figs. 2(b) and 2(c). This morphological evolution occurs due to significant Ehrlich–Schwöebel barriers (also called step-edge barriers) that inhibit movement of atoms between layers of the growing interface.[36–38] Here, the onset of morphological instability does not indicate thermodynamic instability. The lattice remains homogeneous, and the growth continues within the mesoscopic 2.5D growth regime, also known as the SK growth mode.[36,37,39] The InSbBi film will be strained because of the lattice mismatch between the substrate and the film. Hence, the thin film will grow in a layer-by-layer fashion until it reaches a certain critical thickness, beyond which 3D islands or mounds form through the SK transition. This transition marks a shift from two-dimensional (2D) layer-by-layer growth to 3D mound formation, characteristic of the SK growth mode. The effect of the thickness on the mound formation will be further discussed in Sec. D: *Evolution of the mound morphology with thickness*.

In a strained film, the formation of mounds is energetically favorable as it reduces the strain energy in the crystal.[40] The mound density is higher in Fig. 2(c) compared to Fig. 2(b) due to the increased incorporation of Bi into the film with higher Bi flux. As Bi incorporation increases, the lattice constant of the InSbBi thin film increases, leading to a greater lattice mismatch between the substrate and the thin film. Consequently, more mounds with smaller lateral sizes form to reduce the strain energy of the thin film. As the Bi flux increases to $3.3\times10^{-8}$ Torr, the mound morphology in Fig. 2(c) changes, and droplets and elongated nanostructures aligned along the [110] direction begin to form as shown in Fig. 2(d). The presence of these droplets and nanostructures indicates non-substitutional Bi incorporation.[14] As the Bi flux further increases to $5.0\times10^{-8}$ Torr, the nanostructure size and the droplet density increase as shown in Fig. 2(e). The roughness increases exponentially with Bi flux as shown in Fig. 2(f), indicating the existence of a threshold Bi flux between $1.7\times10^{-8}$ and $3.3\times10^{-8}$ Torr for obtaining InSbBi surfaces free of Bi droplets and nanostructures. Large-area 10 × 10 μm$^2$ AFM scans were performed to examine the overall surface morphology over a wider region, as shown in Fig. S1(a-e). The RMS roughness of these scans depicted in Fig. S1(f) exhibits the same trend observed in Fig. 2, where the RMS roughness increases with increasing Bi flux.

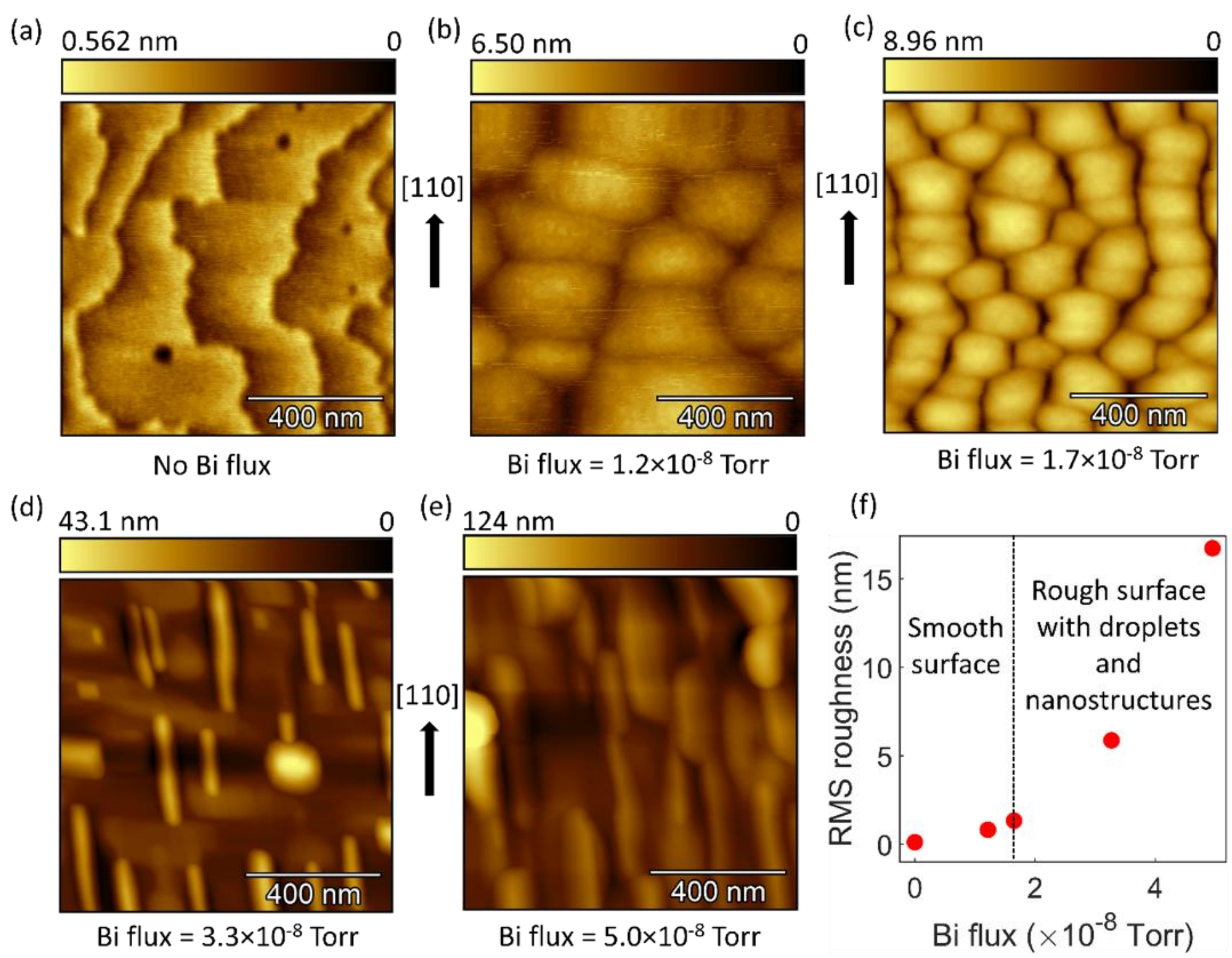

FIG. 2: (a-e) Evolution of surface morphology with at a constant growth temperature of 200°C in 1 × 1 μm$^2$ AFM images at Bi flux = 0 Torr, $1.2\times10^{-8}$ Torr, $1.7\times10^{-8}$ Torr, $3.3\times10^{-8}$ Torr, and $5.0\times10^{-8}$ Torr, respectively. The black arrow corresponds to the [110] direction in the AFM images. (f) RMS roughness calculated from AFM images. The surface is smooth when the RMS roughness is < 2 nm. The surface is rough when the RMS roughness is > 2 nm.

To assess the crystallography and composition of Bi incorporated into the InSbBi thin films, XRD was conducted as shown in Fig. 3(a). An as-received InSb(001) substrate without the InSb buffer layer and the InSbBi film indicated in the blue curve reveals an InSb(004) XRD peak at 56.79° with a shoulder peak at 56.83°, corresponding to the $Sb_2O_4$(222) peak. This $Sb_2O_4$(222) peak originates from the native oxide layer present on the as-received InSb(001) substrate.

This peak is absent in the other curves because the oxide layer in the as-received InSb(001) substrate is removed through thermal desorption in the presence of Sb. While a new oxide layer will be formed upon exposure to the atmosphere after the InSbBi thin film is taken out of the MBE chamber after the completion of the growth, it will be thin and amorphous, lacking the crystalline characteristics required for detectable XRD reflections.

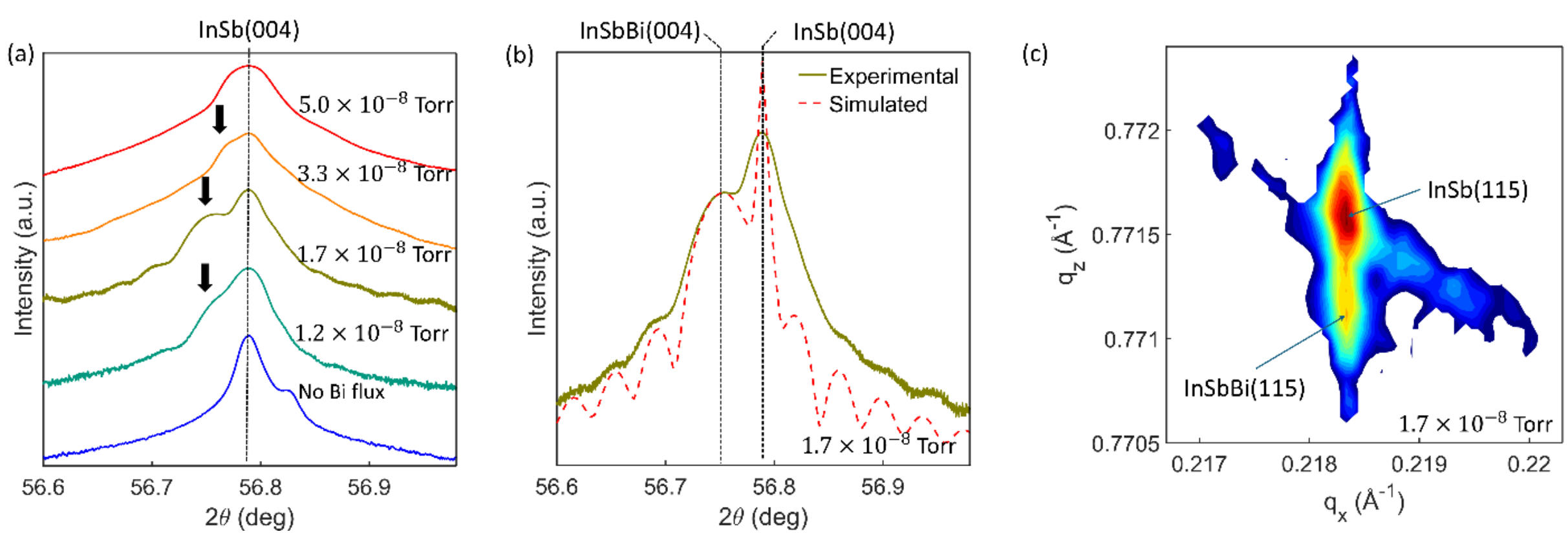

FIG. 3: (a) θ/2θ XRD results of the InSbBi thin films with varying Bi flux from no flux to $5.0\times10^{-8}$ Torr. Black arrows correspond to the InSbBi(004) peaks. (b) Experimental and simulated XRD results for the InSbBi thin film with a Bi flux of $1.7\times10^{-8}$ Torr. (c) The RSM results for the InSbBi thin film with a Bi flux of $1.7\times10^{-8}$ Torr in (b). The first XRD curve in (a) plotted in blue is for the InSb(001) substrate without any growth procedure.

When the Bi flux increases to $1.2\times10^{-8}$ Torr, an InSbBi(004) peak appears, indicating successful Bi incorporation into InSb to form InSbBi. As the Bi flux increases to $1.7\times10^{-8}$ Torr, Bi incorporation further increases, as evidenced by the leftward shift of the InSbBi(400) peak. This shift occurs because the incorporation of Bi expands the InSbBi lattice constant. When the Bi flux is further increased, the InSbBi peak becomes less prominent (orange curve) and nearly vanishes at $5.0\times10^{-8}$ Torr (red curve). Excess Bi flux prevents substitutional incorporation, resulting in Bi accumulation on the surface and the formation of droplets and nanostructures, as indicated by the AFM results for Bi fluxes above $3.3\times10^{-8}$ Torr as shown in Figs. 2(d) and 2(e). Hence, a Bi flux of $1.7\times10^{-8}$ Torr yields the best surface quality and epitaxial InSbBi thin film. The Bi incorporation was estimated by fitting dynamical XRD simulations, as shown in Fig. 3(b). Additionally, Laue oscillations are visible when the Bi flux is $1.7\times10^{-8}$ Torr as illustrated by the experimental XRD curve in Fig. 3(b), indicating high crystallinity and the uniformity of the InSbBi film and smoothness of the interface between the InSbBi and InSb[41-43]. When the Bi flux is $1.2\times10^{-8}$ Torr, the Bi incorporation is 0.9%, and when the Bi flux increases to $1.7\times10^{-8}$ Torr, the Bi incorporation is 1.81%. Therefore, by increasing the Bi flux up to a certain threshold ($3.3\times10^{-8}$ Torr), the Bi incorporation in the InSbBi film can be tuned.

To gain a comprehensive understanding of the strain in InSbBi thin films grown on an InSb(001) substrate, a reciprocal space map (RSM) was performed on the sample grown with a Bi flux of $1.7\times10^{-8}$ Torr, focusing on the (115) reflection, as illustrated in Fig. 3(c). The corresponding XRD data is presented in Fig. 3(b). The presence of strain in the InSbBi thin film is evident from the alignment of the InSbBi(115) peak with the InSb(115) peak along $q_x$, indicating that the in-plane lattice parameter of the film matches that of the substrate.

*Table 1: Calculation of strain on InSbBi thin films obtained from RSM results.*

| Out-of-plane lattice constant (XRD) [Å] | In-plane lattice constant (RSM) [Å] | Out-of-plane lattice constant (RSM) [Å] | In-plane strain [%] | Out-of-plane strain [%] | Poisson ratio |
|---|---|---|---|---|---|
| 6.483 | 6.477 | 6.484 | 0.093 | 0.015 | 0.087 |

The in-plane and out-of-plane strains were quantified using lattice constants extracted from the RSM peaks, as summarized in Table 1. The strain ($\varepsilon$) was calculated using the relation $\varepsilon = (a_{strained} -$ $a_0)/a_0$, where $a_{strained}$ represents either the in-plane or out-of-plane lattice parameter of the strained InSbBi film, and $a_0$ is the relaxed lattice parameter of InSbBi (6.483 Å). The relaxed lattice constant was calculated from the Vegard's law combined with a bowing coefficient using $a_{\mathrm{InSbBi}} = xa_{\mathrm{InBi}} + (1-x)a_{\mathrm{InSb}} + 0.0012x(1-x)$, where $a_{\mathrm{InSbBi}}$, $a_{\mathrm{InBi}}$, $a_{\mathrm{InSb}}$ are the lattice constants of $InSb_{1-x}Bi_x$, InBi, and InSb, respectively[44]. The Poisson ratio (ν) was determined using the equation

$\nu = \frac{\varepsilon_{\mathrm{out-of-plane}}}{\varepsilon_{\mathrm{out-of-plane}} - 2\times\varepsilon_{\mathrm{in-plane}}}$, where $\varepsilon_{\mathrm{out-of-plane}}$ and $\varepsilon_{\mathrm{in-plane}}$ represents the out-of-plane and in-plane strains, respectively. [45,46] Here, the out-of-plane strain is multiplied by -1 to account for the tensile in-plane strain of the InSbBi film.[45,46] Notably, the out-of-plane lattice constants derived from both RSM and XRD measurements exhibit close agreement, further validating the strain analysis in the InSbBi thin films.

## C. Change in surface morphology with varying growth temperature

The effect of growth temperature on the morphology of InSbBi thin films was investigated by varying the temperature from 183°C to 250°C while maintaining a constant Bi flux of $1.7\times10^{-8}$ Torr. A significant surface morphology change occurs with increasing temperature, as shown in Fig. 4. At a growth temperature of 183°C, the surface shows a rough morphology containing nanostructures and small droplets possibly due to the deposition of Bi on the surface as illustrated in Fig. 4(a). When the growth temperature is increased to 200°C, the InSbBi thin films show the characteristic surface morphology of InSbBi with mounds, as shown in Fig. 4(b). This is the same sample that was discussed earlier in Fig. 2(c). With increasing temperature, the mound morphology of the sample changes. At 217°C , the smaller mounds shown in Fig. 4(b) coalesce to create larger mounds, while the height of the mounds decreases, as shown in Fig. 4(c). At 233°C, the onset of the step edges occurs as shown in Fig. 4(d). This indicates a significant shift from the mound morphology observed for 200°C and 217°C to the step-flow growth with step meandering.[36,37] At 250°C, the step-flow growth mode is more pronounced, as shown in Fig. 4(e). The reason for this behavior is that at elevated temperatures the energetic adatoms have higher mobility, allowing them to migrate across the surface and reach steps for attachment and step formation[47]. The size of the mounds is approximately 200 × 200 nm$^2$ (Fig. 4(f)). The 3D topography of the step-flow growth is shown in Fig. 4(g). The average height of a step is approximately (0.324 ± 0.001) nm, as illustrated in Fig. 4(h), which is close to the monolayer height of InSb (0.324 nm)[48] which further indicates that the percentage incorporation of Bi is close to zero at 250°C.

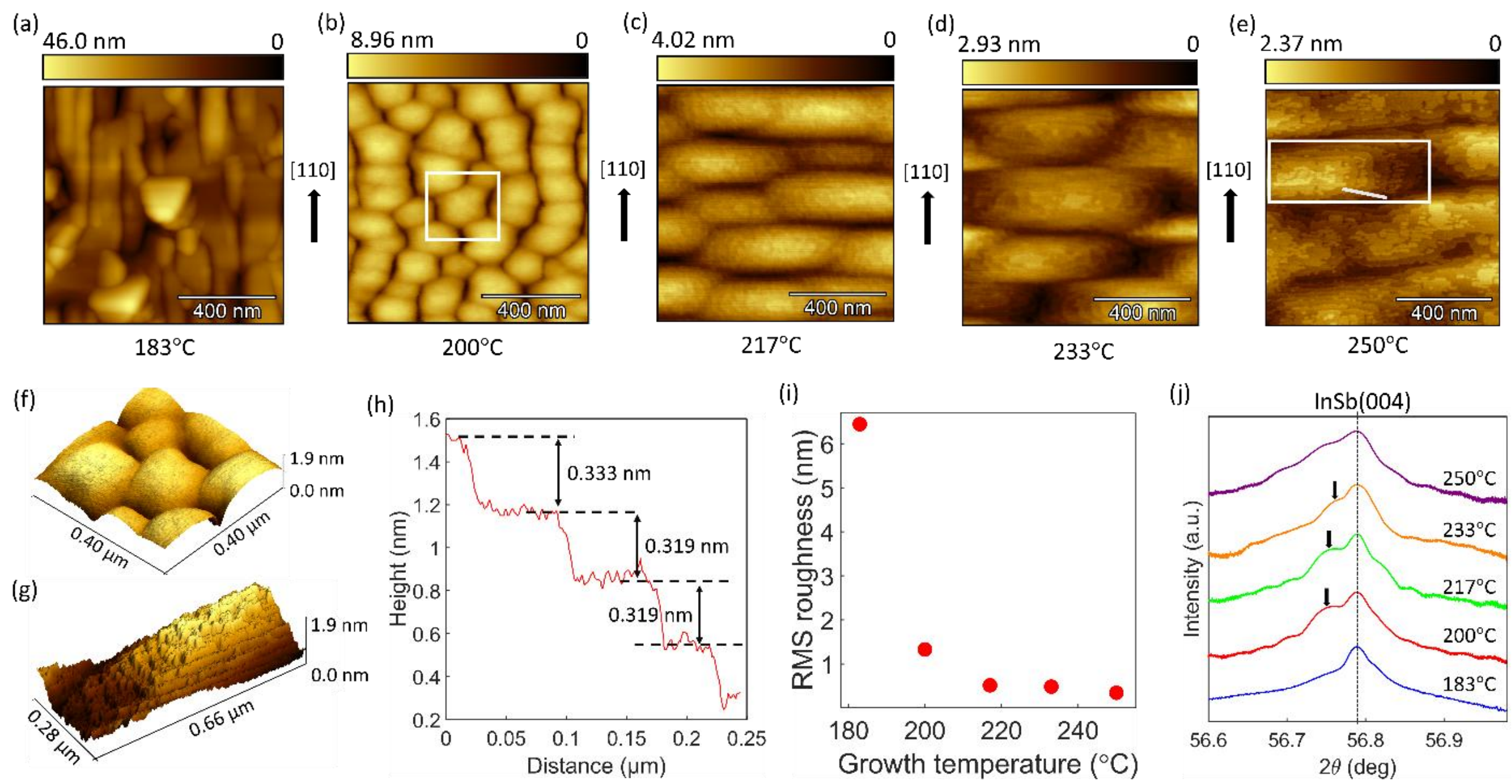

FIG. 4: (a-e) Change in surface morphology with the growth temperature at a constant Bi flux of 1.7×10$^{-8}$ Torr are shown in 1×1 μm$^2$ AFM images at growth temperature = 183°C, 200°C, 217°C, 233°C, and 250°C, respectively. The black arrows directed up correspond to the [110] direction. (f,g) 3D topographies of the areas indicated in the white rectangles in (b) and (e) representing the mound morphology and the step-flow growth respectively. (h) Line profile of the AFM image in (e) along the white line. (i) RMS roughness obtained from the AFM images. (j) $\theta/2\theta$ XRD results for the InSbBi thin films corresponding to the samples indicated in the AFM images. The black arrows directed down correspond to the InSbBi(004) peaks.

With increasing temperature, the surface gets smoother as shown by the RMS roughness decreasing exponentially (Fig. 4(i)), which happens because the surface changes from having a mound morphology to that of step-flow growth. This can be attributed to the fact that the probability of Bi incorporating into the film to form InSbBi at higher temperatures will decrease owing to the large size of Bi atoms.[49] In fact, this has been observed in GaSbBi and GaAsBi films where Bi has been incorporated to form III–V alloys.[49-51]

To evaluate the surface morphology across a broader area, large-area 10×10 $\mu m^2$ AFM scans were performed, as shown in Fig. S2(a–e). The corresponding RMS roughness values, presented in Fig. S2(f), reveal a non-monotonic trend: the roughness initially decreases as the growth temperature increases from 183°C to 217°C, followed by a pronounced exponential increase from 217°C to 250°C. This behavior contrasts with the trend observed in Fig. 4(i), where the RMS roughness decreases exponentially with increasing temperature.

The increase in roughness at higher temperatures may be attributed to enhanced surface mobility of Bi atoms, which increases the likelihood of Bi atom clustering and subsequent droplet formation. Hence, it can promote droplet formation. This is further supported by the fact that the Bi–Bi bond energy (200 kJ/mol) is significantly higher than that of In–Bi (153.6 kJ/mol).[52] Hence, this makes Bi droplet formation energetically more favorable than Bi incorporation into substitutional sites within the InSbBi lattice.[1] The interplay of these factors likely contributes to the observed surface roughening at elevated temperatures. These findings highlight a clear evolution in surface morphology over the temperature range from 183°C to 250°C, emphasizing the critical role of growth temperature in achieving smooth InSbBi thin films.

To assess the crystallography and the percentage of the Bi incorporated into the InSbBi thin films, XRD was conducted as shown in Fig. 4(j). When the growth temperature is 200°C, there is a clear InSbBi(400) peak. As the growth temperature increases, the peak shifts right towards the InSb(004) substrate peak. This is to be expected as the Bi incorporation to form InSbBi decreases with the increase in temperature. The lattice constant decreases with the decrease of Bi percentage in InSbBi, which is reflected as a shift in the XRD peak corresponding to InSbBi. The XRD plot which corresponds to 250°C, which is indicated in red in Fig. 4(j), has no discernible InSbBi(004) peak because the Bi incorporation exponentially decreases at elevated growth temperatures. Hence, at a growth temperature of 250°C, Bi isn't incorporated to form InSbBi. The presence of Laue oscillations in the plots corresponding to 200°C, 217°C, and 233°C in Fig. 4(j) indicates the high crystallinity and the uniformity of the InSbBi film and smoothness of the interface between the InSbBi and InSb.[41-43] The presence of Laue oscillations in the plot corresponding to 250°C in Fig. 4(j) indicates the high crystallinity and the uniformity of the InSb film when Bi is not incorporated to form InSbBi thin film.[41-43]

### D. Evolution of the mound morphology with thickness

The evolution of the mound morphology was studied by systematically varying the InSbBi film thickness while maintaining a constant Bi flux of $1.7\times10^{-8}$ Torr and the growth temperature of 200°C (Fig. 4). InSb exhibits a step-edge morphology resulting from the step-flow growth mode, as shown in Fig. 5(a). When 20 nm of InSbBi is grown, the step-flow morphology of InSb becomes less prominent, and the onset of mound formation is clearly visible on the terraces. As the InSbBi film thickness increases, small mounds coalesce and grow. At a thickness of 150 nm, the tops of the mounds develop additional smaller mounds, as shown in Fig. 5(c). This occurs because arriving adatoms preferentially nucleate on top of existing mounds, resulting in the sequential formation of second-layer nuclei, followed by third-layer nucleation, and so on. When the film reaches 300 nm in thickness, the mounds appear smoother (Fig. 5(d)). This 300-nm-thick InSbBi is the same sample presented in Figs. 2(c) and 4(b). The primary reason for the smoothing is strain relaxation with increasing thickness. As the InSbBi film grows thicker, strain gradually relaxes, reducing surface instabilities that contribute to roughening. This is further supported by the minimal in-plane and out-of-plane strains observed (Table 1). Although the film becomes smoother in appearance, the RMS roughness increases with thickness (Fig. 5(d)), primarily due to the increasing mound height.

XRD scans reveal the crystallographic structure and the composition of the films, as shown in Fig. 5(f). InSbBi(004) peaks are clearly visible, with the peaks shifting towards the InSb(004) peak as the film thickness increases. This shift can be attributed to the film relaxation that occurs with increased thickness, allowing for a reduction in strain and closer alignment with the InSb(004) substrate peak. Specifically, the out-of-plane strains calculated from XRD for the 20 nm and 150 nm InSbBi films are 0.031% and 0.021%, respectively. These values are higher than the out-of-plane strain observed for the 300 nm sample, as shown in Table 1. The presence of Laue oscillations in the samples containing InSbBi thin films indicates the high crystallinity and the uniformity of the InSbBi thin film.[41-43]

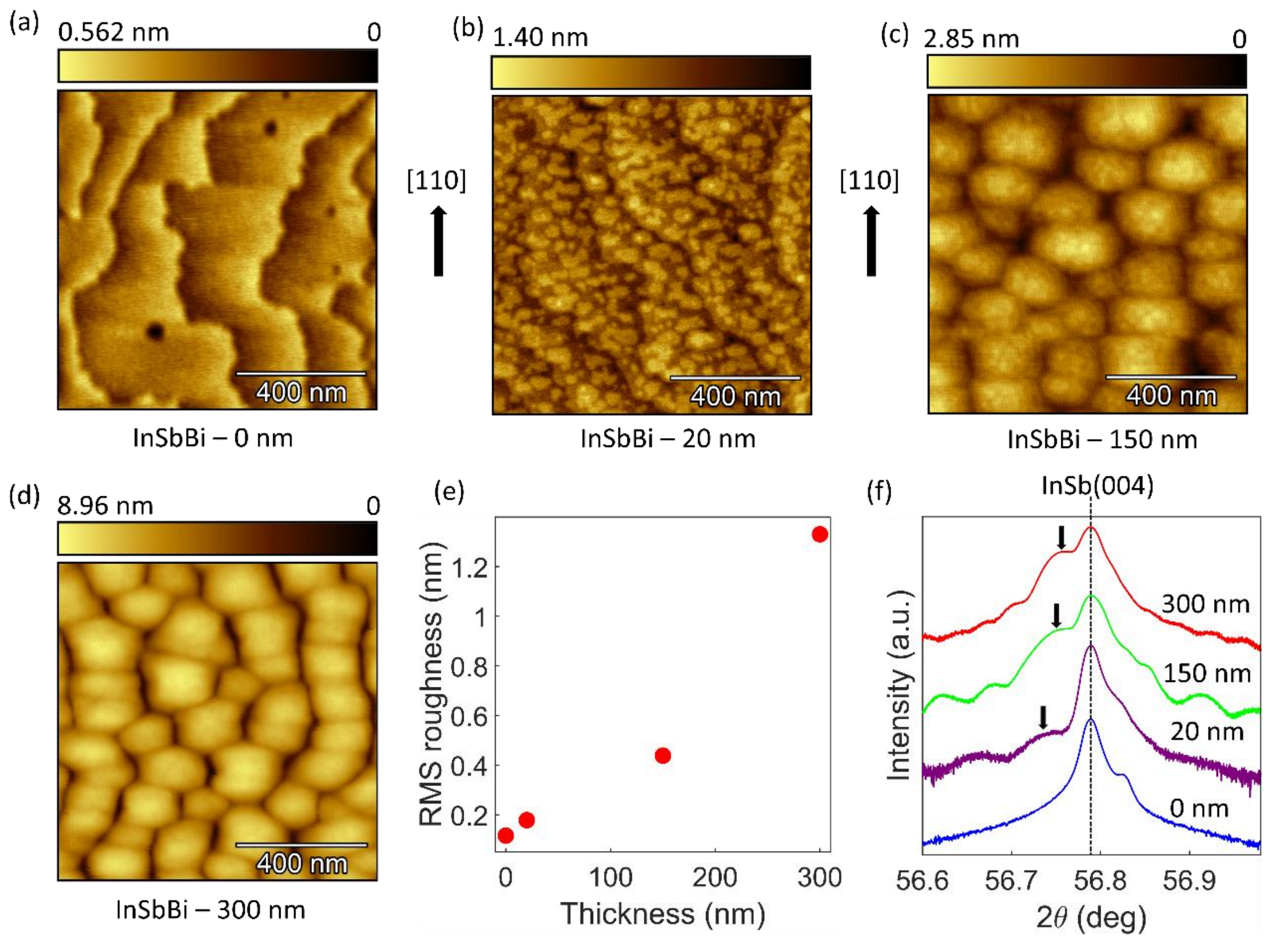

FIG. 5: (a-d) Change in surface morphology with thickness at a constant growth temperature of 200°C at a constant Bi flux of $1.7\times10^{-8}$ Torr shown in $1\times1$ $\mu m^2$ AFM images at thickness 0 nm, 20 nm, 150

nm and 300 nm, respectively. (e) RMS roughness calculated from the AFM images. (f) θ/2θ XRD results for the InSbBi thin films corresponding to the samples indicated in the AFM images. The black arrows directed down correspond to the InSbBi(004) peaks.

Hence, it is evident that the Bi incorporation can be tuned by either changing the Bi flux (Fig. 6(a)) or the growth temperature (Fig. 6(b)), keeping the other growth parameters constant. A maximum of 1.8 % Bi incorporation was achieved having smooth surface morphology with an RMS roughness < 2 nm.

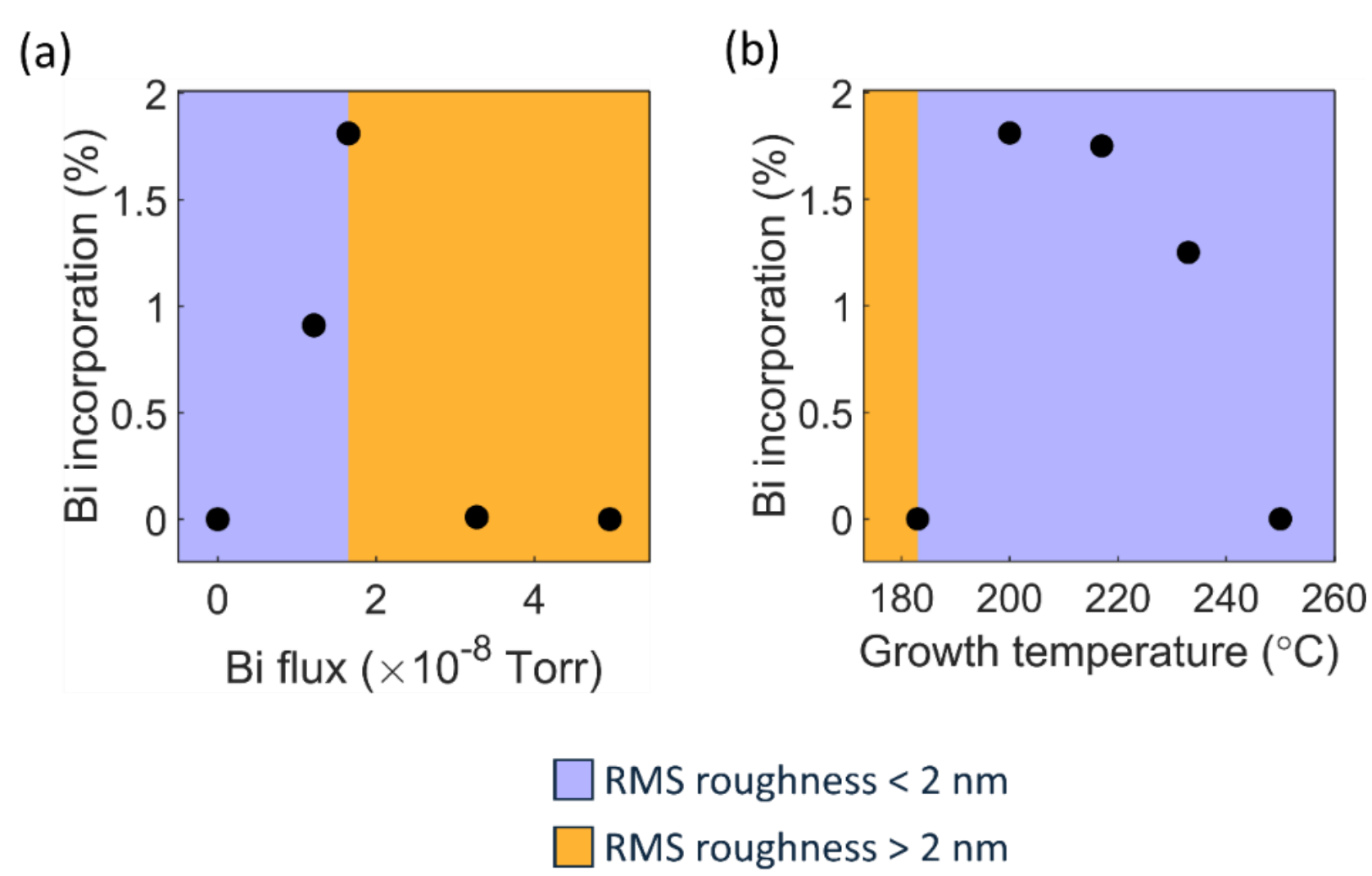

FIG. 6 : Bi incorporation as a function of (a) Bi flux at a constant growth temperature of 200°C, and (b) growth temperature at a fixed Bi flux of 1.7×10$^{-8}$ Torr. The purple and yellow regions denote the parameter windows corresponding to smooth (RMS roughness < 2 nm) and rough (RMS roughness > 2 nm) surface morphologies, respectively. Data points located at the boundary between the two regions correspond to samples with RMS roughness below 2 nm.

## IV. CONCLUSION

In summary, we have established a robust methodology for growing high-quality InSbBi thin films with Bi incorporation of up to 1.81% on InSb(001). Our findings reveal a distinct transition in growth mode, shifting from a mound-dominated morphology to step-flow growth, driven by the interplay between growth parameters and Bi incorporation. Notably, smooth thin films with mound surface morphology exhibit RMS roughness below 2 nm, quantified by AFM. Precise Bi incorporation is determined by fitting simulated and experimental XRD results. We demonstrate that achieving atomically smooth films requires maintaining a V/III flux ratio of unity, dynamically calibrated via RHEED analysis. Furthermore, Bi incorporation can be finely tuned by varying the growth temperature between 200°C and 250°C or adjusting the Bi flux between $1.2\times10^{-8}$ and $1.7\times10^{-8}$ Torr while keeping other growth parameters constant. These results mark a significant advancement in the growth of high-quality InSbBi thin films, laying the foundation for their integration in the next-generation optoelectronic and quantum device applications.

## SUPPLEMENTARY MATERIAL

Large-area 10 × 10 $\mu m^2$ AFM images corresponding to variations in Bi flux at a fixed growth temperature, as well as changes in growth temperature at a constant Bi flux.

## ACKNOWLEDGMENTS

This work was supported by the Science Alliance at the University of Tennessee, Knoxville, through the Support for Affiliated Research Teams (StART) program. P.A. acknowledges support from the U.S. Department of Energy, Office of Science, Basic Energy Sciences, Materials Sciences and Engineering Division for the material growth efforts. Structural characterizations were carried out as part of a user project at the Center for Nanophase Materials Sciences, which is a U.S. Department of Energy, Office of Science User Facility at Oak Ridge National Laboratory.

## AUTHOR DECLARATIONS

The authors have no conflicts to disclose.

## DATA AVAILABILITY

The data that supports the findings of this study are available from the corresponding author upon reasonable request.

# Evolution of surface morphology from Stranski–Krastanov growth mode to step-flow growth mode in InSbBi thin films

Chandima Kasun Edirisinghe[1], Anuradha Wijesinghe[1], Anjali Rathore[1], Pradip Adhikari[1], Christopher Rouleau[2], Joon Sue Lee[1,*]

[1]Department of Physics and Astronomy, University of Tennessee, Knoxville, TN 37996, USA

[2]Center for Nanophase Materials Sciences division, Oak Ridge National Laboratory, USA

[*]Corresponding author e-mail: jslee@utk.edu

SUPPLEMENTARY MATERIAL

### A. AFM images when the Bi flux is changed keeping the growth temperature constant

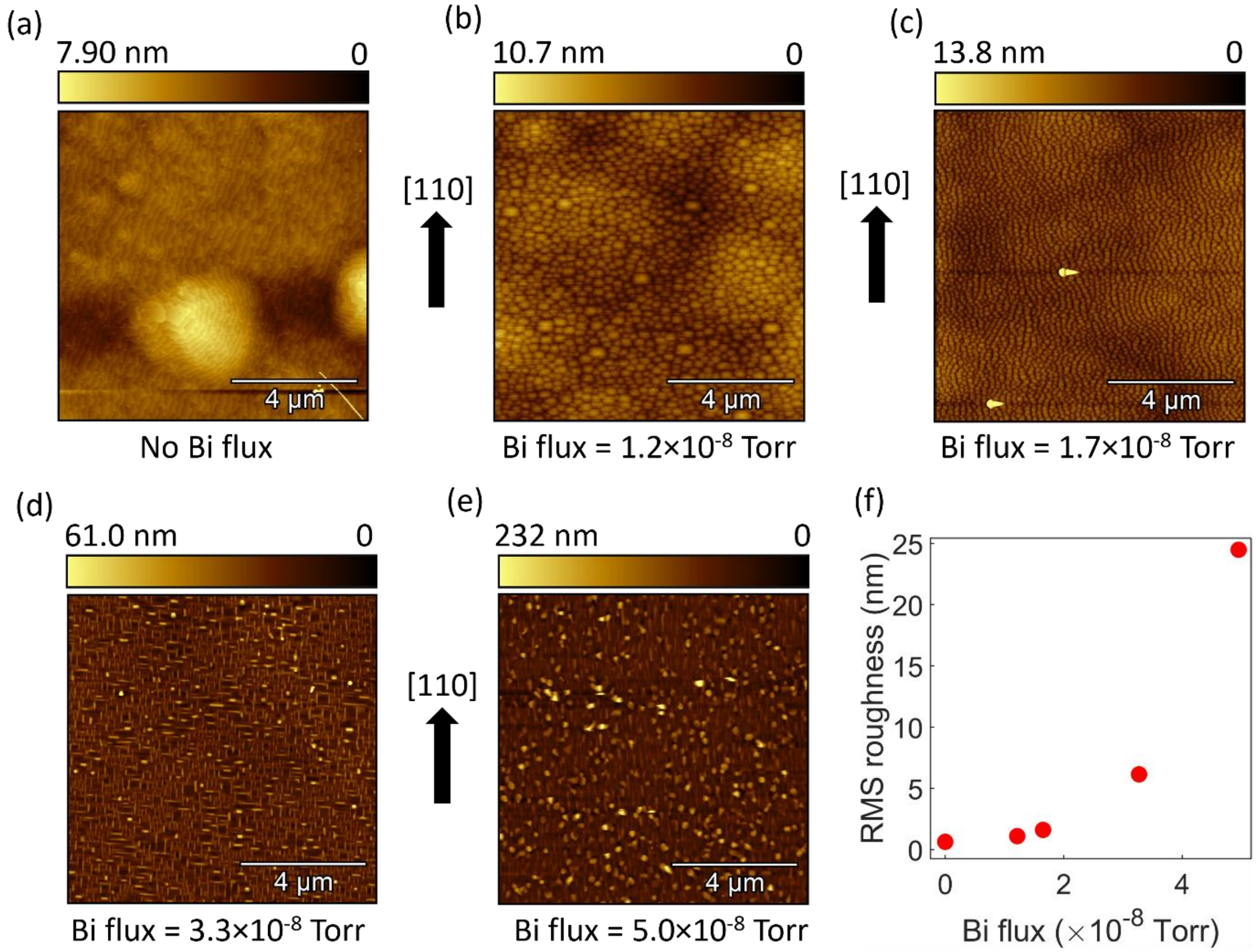

FIG. S7: (a-e) Evolution of surface morphology with at a constant growth temperature of 200°C in 10 × 10 μm$^2$ AFM scans at Bi flux = 0 Torr, $1.2\times10^{-8}$ Torr, $1.7\times10^{-8}$ Torr, $3.3\times10^{-8}$ Torr, and $5.0\times10^{-8}$ Torr respectively. The black arrow corresponds to the [110] direction in the AFM images. (f) RMS roughness calculated from AFM images.

**B. AFM images when the temperaturelux is changed keeping the growth temperature constant**

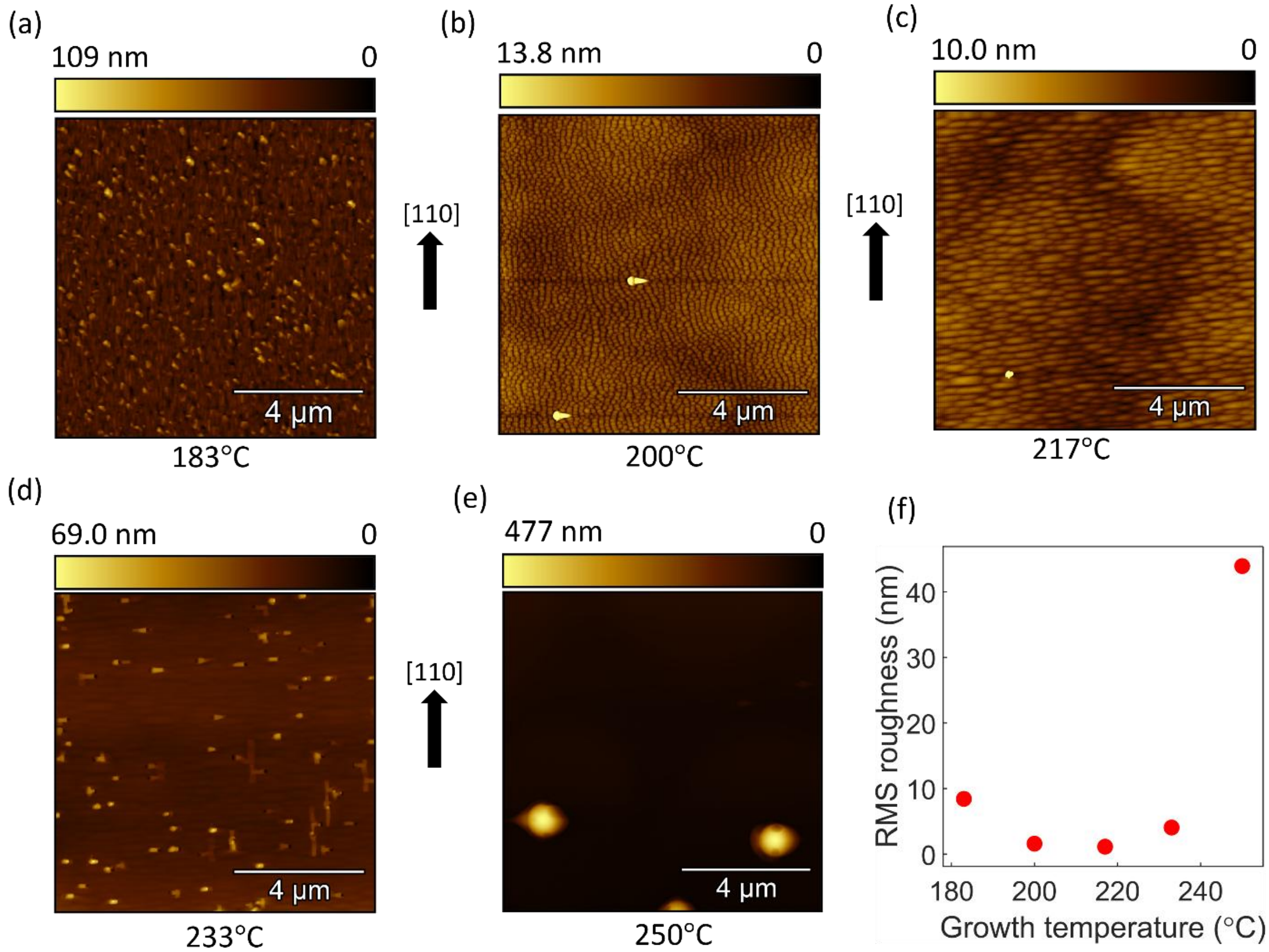

FIG. S2: (a-e) Change in surface morphology with the growth temperature at a constant Bi flux of $1.7\times10^{-8}$ Torr are shown in $10\times10$ $\mu m^2$ AFM images at growth temperature = 183ºC, 200°C, 217°C, 233°C, and 250°C respectively. The black arrows correspond to the [110] direction. (f) RMS roughness calculated from AFM images.